\documentstyle[aps,twocolumn,epsf]{revtex}

\def\psfig#1#2{ \begin{center}
                   \epsfxsize=#2
                    \leavevmode\epsffile{#1.EPS}
                \end{center} }

\title{Entanglement of coherent states and decoherence}
\author{Radim Filip, Jaroslav
\v{R}eh\'{a}\v{c}ek and Miloslav Du\v{s}ek}
\address{Department of Optics, Palack\'{y} University,
17.~listopadu 50, 772\,00 Olomouc, Czech Republic}

\begin{document}
\maketitle
\begin{abstract}
A possibility to produce entangled superpositions of strong coherent
states is discussed. A recent proposal by Howell and Yazell [{\em Phys.
Rev. A 62, 012102 (2000) \/}] of a device which entangles two strong
coherent coherent states is critically examined. A serious flaw in
their design is found. New modified scheme is proposed and it is shown
that it really can generate non-classical states that can violate Bell
inequality. Moreover, a profound analysis of the effect of losses and
decoherence on the degree of entanglement is accomplished. It reveals
the high sensitivity of the device to any disturbances and the fragility
of generated states.
\end{abstract}

\narrowtext

\section{Introduction}

Quite recently an interesting idea has been proposed \cite{howell}
how to entangle two strong coherent fields and generate
a four-mode Schr\"{o}dinger cat-like state utilizing
quantum non-demolition (QND) measurement.
The idea has been inspired by recent works
on a similar subject \cite{macro}.
However, the authors
made a serious mistake in their reasoning, which renders
their experimental scheme unworkable.
Before we show how to modify the scheme in order to
get the desired effect we shall briefly discuss the
original experimental arrangement.

The original scheme for entangling ``macroscopic'' fields is
that outlined in Fig.~\ref{schema} without beam splitter BS
and detectors D$_1$ and D$_2$. The idea
goes as follows. First, a non-separable single photon state
%%%%%%%%%%%%%%%%%%%%%%%%%%%%%%%%%%%%%%%%%%%%%%%%%
\begin{equation} \label{single}
(|1\rangle_{12}|0\rangle_{13}+i|0\rangle_{12}|1\rangle_{13})/\sqrt{2}
\end{equation}
%%%%%%%%%%%%%%%%%%%%%%%%%%%%%%%%%%%%%%%%%%%%%%%%%
is produced using the single-photon source SPS and a 50:50
beam splitter.
In the next step an attempt is made to transfer the which-path
uncertainty of the photon into the entanglement of strong
coherent fields generated by the coherent sources CS by means
of non-demolition measurements of the number of photons
in modes $12$ and $13$. The QND measuring devices
operate by the cross-Kerr interaction described by the
Hamiltonian
%%%%%%%%%%%%%%%%%%%%%%%%%%%%%%%%%%%%%%%%%%%%%%%%%
\begin{equation} \label{hamilt}
H_{\rm QND}= \hbar\pi (\hat{n}_{12}\hat{n}_{23} +
              \hat{n}_{13}\hat{n}_{33}),
\end{equation}
%%%%%%%%%%%%%%%%%%%%%%%%%%%%%%%%%%%%%%%%%%%%%%%%%
where $\hat{n}$ are photon number operators of the corresponding modes.
The strengths of the interactions (\ref{hamilt}) are carefully
chosen so as to yield the accumulated phase shift of $\pi$
in modes $23$ or $33$ provided a single photon is present
in modes $12$ or $13$, respectively.
This means that the presence of a single photon, for instance, in
$12$ mode will cause the outputs of $24$ and $25$ modes
to be switched completely. This switching lies in the heart of
the quantum entangling device which according to Howell and Yeazell
\cite{howell} should generate four-mode ``macroscopic''
entangled state
%%%%%%%%%%%%%%%%%%%%%%%%%%%%%%%%%%%%%%%%%%%%%%%%%%%%
\begin{equation} \label{puvodni}
|\Phi\rangle=N_+(|\phi_1\rangle+|\phi_2\rangle),
\end{equation}
%%%%%%%%%%%%%%%%%%%%%%%%%%%%%%%%%%%%%%%%%%%%%%%%%%%%
where
%%%%%%%%%%%%%%%%%%%%%%%%%%%%%%%%%%%%%%%%%%%%%%%%%%%%%
\begin{eqnarray} \label{def}
|\phi_1\rangle&=&|\alpha_2\rangle_{24}|0\rangle_{25}
|0\rangle_{34}|\alpha_3\rangle_{35} \nonumber \\
|\phi_2\rangle&=&|0\rangle_{24}|\alpha_2\rangle_{25}
|\alpha_3\rangle_{34}|0\rangle_{35}.
\end{eqnarray}
%%%%%%%%%%%%%%%%%%%%%%%%%%%%%%%%%%%%%%%%%%%%%%%%%%%%%
Here $\alpha_2$ and $\alpha_3$ are complex amplitudes
proportional to the amplitudes of the coherent fields
fed into the inputs $21$ and $31$, respectively, and
$N_+$ is a normalization factor \cite{normalizace}.
However, the four-mode output state was not calculated
explicitly in \cite{howell}. The authors rather guessed
at the four-mode state from the results of a few
thought experiments with the output light.
The authors' claim that the observed correlations
are of a purely quantum nature and could not be explained
classically is in error, though.
It could be easily demonstrated that the correlations
discussed in \cite{howell} can be explained
by the following mixed four-mode state,
%%%%%%%%%%%%%%%%%%%%%%%%%%%%%%%%%%%%%%%%%%%%%%%%%%%%%
\begin{equation} \label{classic}
\rho=\frac{1}{2}(|\phi_1\rangle\langle\phi_1|+
     |\phi_2\rangle\langle\phi_2|),
\end{equation}
%%%%%%%%%%%%%%%%%%%%%%%%%%%%%%%%%%%%%%%%%%%%%%%%%%%%%
that is by the ``classical'' statistical mixture,
which is usually referred to as
a separable mixture or a state containing only
classical correlations. It is not difficult to see that
it is the mixed (classical) state (\ref{classic}) rather
than cat state (\ref{puvodni}) what is produced
by the original Howell and Yeazell's apparatus.
Working in the Schr\"{o}dinger picture, the
output  six-mode state can straightforwardly be
calculated as follows
%%%%%%%%%%%%%%%%%%%%%%%%%%%%%%%%%%%%%%%%%%%%%%%%%%%%%
\begin{equation} \label{cely}
|\Phi_{\rm all}\rangle=\frac{1}{\sqrt{2}}\left(|\phi_1\rangle|0\rangle_{12}|1\rangle_{13}+
|\phi_2\rangle|1\rangle_{12}|0\rangle_{13}\right).
\end{equation}
%%%%%%%%%%%%%%%%%%%%%%%%%%%%%%%%%%%%%%%%%%%%%%%%%%%%%
It is the photon leaving the apparatus in modes $12$, $13$ and
carrying which-way information what destroys
the entanglement among the remaining four modes.
Mathematically, the output state of four modes
$24$, $25$, $34$, $35$ is obtained by tracing the density matrix
$|\Phi_{\rm all}\rangle\langle\Phi_{\rm all}|$ over the two modes;
we end up with the mixed state  (\ref{classic}).
This somehow slipped attention of the authors
in \cite{howell}.

\section{QND entangling device}

\begin{figure}
\psfig{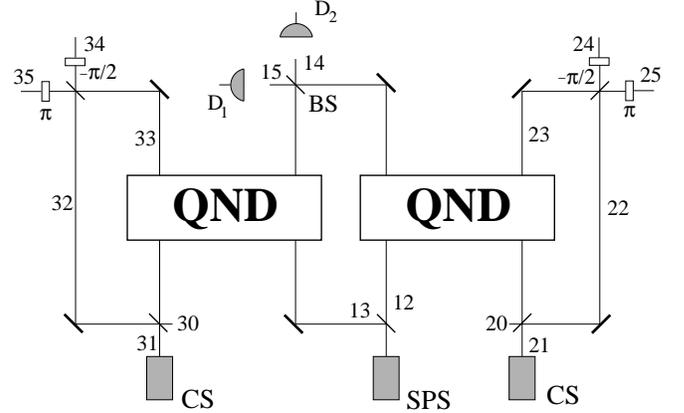}{\hsize}
\caption{Entangling apparatus. QND -- quantum non-demolition
measuring devices; CS -- coherent state sources; SPS -- single-photon
source; BS -- beam splitter; D$_1$, D$_2$ -- single-photon detectors.}
\label{schema}
\end{figure}

Having identified the catch in the original scheme
one can try to work around it. The key point
is to erase the which-way information that resides
in $12$ and $13$ two-mode field when the interaction is over.
To accomplish this we propose to superimpose these modes
at a 50:50 beam splitter BS as shown in Fig.~\ref{schema}.
Two single photon detectors D$_1$ and D$_2$ are attached
to the outputs of beam splitter BS.
After mixing the $12$ and $13$ beams at the beam splitter the state
of the six-mode field reads
%%%%%%%%%%%%%%%%%%%%%%%%%%%%%%%%%%%%%%%%%%%%%%%%%%%%%%%%%%%
\begin{eqnarray} \label{mix}
|\Phi_{\rm all}\rangle&=&
\frac{1}{2}\left[-(|\phi_1\rangle-|\phi_2\rangle)
  |0\rangle_{14}|1\rangle_{15}\right. \nonumber \\
  &&+i\left.(|\phi_1\rangle+|\phi_2\rangle)|1\rangle_{14}|0\rangle_{15}
\right].
\end{eqnarray}
%%%%%%%%%%%%%%%%%%%%%%%%%%%%%%%%%%%%%%%%%%%%%%%%%%%%%%%%%%%%
Now, depending on the result of the single-photon detection,
the four-mode state of interest becomes
%%%%%%%%%%%%%%%%%%%%%%%%%%%%%%%%%%%%%%%%%%%%%%%%%%%%%%%%
\begin{equation} \label{bily}
|\Phi_1\rangle=N_+(|\phi_1\rangle+|\phi_2\rangle),
\end{equation}
%%%%%%%%%%%%%%%%%%%%%%%%%%%%%%%%%%%%%%%%%%%%%%%%%%%%%%
if D$_1$ detector fires, or
%%%%%%%%%%%%%%%%%%%%%%%%%%%%%%%%%%%%%%%%%%%%%%%%%%%%%%
\begin{equation} \label{cerny}
|\Phi_2\rangle=N_-(|\phi_1\rangle-|\phi_2\rangle),
\end{equation}
%%%%%%%%%%%%%%%%%%%%%%%%%%%%%%%%%%%%%%%%%%%%%%%%%%%%%%
if D$_2$ does. $N_-$ is a normalization factor \cite{normalizace}.
We remind that for
$|\alpha_2|,|\alpha_3|\gg1$ the four-mode states $|\phi_1\rangle$ and
$|\phi_2\rangle$ are well separated ``macroscopic'' states.
Hence we arrived at the sought after effect. In fact two
kinds of phase shifted Schr\"{o}dinger cat-like states can
be generated using the scheme depicted in Fig.~\ref{schema}.
Now a perfect source of the ``macroscopic'' entangled field
can be realized using a gate triggered by the D$_1$ and D$_2$
detectors so that only one of the states (\ref{bily}) and (\ref{cerny})
is allowed to go through.

Here a comment seems to be in order. Had the information gained
by means of detectors D$_1$ and D$_2$ been not used for post-selecting
the state of the remaining four modes, we would have had the same
four-mode output field as Howell and Yeazell had.
Ignoring the result of the single-photon detection amounts
to taking the weighted average of the states
$|\Phi_1\rangle$ and $|\Phi_2\rangle$.
It is not difficult to see that the averaging bring us
back to the mixed state (\ref{classic}).
This is consistent with quantum theory because no tampering
with modes $14$, $15$ can influence the results of
experiments performed on the remaining (possibly space-like separated)
beams. We should also mention that the idea used in this
report to correct the original faulty scheme is in fact an
implementation of the well-known and much discussed
quantum eraser \cite{eraser}.

Let us close this section observing that actually
the amended setup (Fig.~\ref{schema}) is unnecessarily too complicated;
it contains redundant parts.
One can easily do with just one QND device.
Let us remove the rightmost  Mach-Zehnder (MZ) interferometer and its QND
device from the setup. The
output state will change to
$(|0\rangle_{34}|\alpha_3\rangle_{35}+
|\alpha_3\rangle_{34}|0\rangle_{35})/\sqrt{2}$,
supposing that a click at detector $D_1$ have been registered.
After using two additional beam splitters in $34$ and $35$ paths
and re-labeling the output modes, the four-mode entangled state
equivalent to (\ref{puvodni}) is obtained again.

\section{Entanglement and nonlocality}

Before discussing decoherence which could be serious obstacle
in the way of realizing the ``macroscopic'' entangler in laboratory,
let us discuss briefly some interesting properties of the
``cat'' state (\ref{puvodni}). First of all one may ask, to which
extend the state (\ref{puvodni}) is entangled.
We will adopt mutual information as a convenient
measure of entanglement \cite{barnett}.
It is defined by
the difference between the sum of von Neumann entropies $S_1$ and $S_2$ of
subsystems $1$ and $2$ and the von Neumann entropy of
the composite system $S$:
%%%%%%%%%%%%%%%%%%%%%%%%%%%%%%%%%%%%%%%%%%%%%%%%%%%%
\begin{equation} \label{entrop}
I=S_1+S_2-S.
\end{equation}
%%%%%%%%%%%%%%%%%%%%%%%%%%%%%%%%%%%%%%%%%%%%%%%%%%%%
Here subsystem $1$ consists of modes $24$ and $25$; subsystem $2$
consists of modes $34$ and $35$. Mutual information (\ref{entrop})
is zero if the subsystems 1 and 2 are uncorrelated
($\rho$=$\rho_1\otimes\rho_2$). Its maximum
value still explainable by classical correlations
is $I=S$ \cite{correl}.

For simplicity we will consider symmetric inputs
$\alpha_2=\alpha_3=\alpha$.
Although explicit calculation of mutual information
$I$ for general input coherent state is tedious,
it is not difficult to ascertain its lower and upper band.
For very small input amplitude $|\alpha|\ll 1$ state
(\ref{puvodni}) becomes nearly a product state, hence its mutual
information
goes to zero. Maximum is attained for very large $|\alpha|$ --
such that the two components in (\ref{puvodni}) become
almost orthogonal states. Then one obtains
%%%%%%%%%%%%%%%%%%%%%%%%%%%%%%%%%%%%%%%%%%%%%%%%%%%%
\begin{equation} \label{nase-entrop2}
I_{\rm max}=2\ln2.
\end{equation}
%%%%%%%%%%%%%%%%%%%%%%%%%%%%%%%%%%%%%%%%%%%%%%%%%%%%

An issue closely related to entanglement is
question whether particular entangled states can violate
local realism \cite{EPR}.
Noticing that for $\alpha\gg1$ two-dimensional subspace spanned
by vectors $|\alpha, 0\rangle_{1}\equiv|\alpha\rangle_{24}|0\rangle_{25}$
and $|0,\alpha\rangle_{1}$, living in Hilbert
space of the first system is isomorphic to spin-half
particle space (similarly for system $2$), one can
define unitary operations
%%%%%%%%%%%%%%%%%%%%%%%%%%%%%%%%%%%%%%%%%%%%%%%%%%%%
\begin{eqnarray} \label{operace}
|\alpha, 0\rangle_{1}&\rightarrow&M_+(\cos\theta\, |\alpha, 0\rangle_{1}
+\sin\theta\, |0,\alpha\rangle_{1}) \nonumber \\
|0,\alpha\rangle_{1}&\rightarrow&
M_-(-\sin\theta\, |\alpha, 0\rangle_{1}+
\cos\theta\, |0, \alpha\rangle_{1})
\end{eqnarray}
%%%%%%%%%%%%%%%%%%%%%%%%%%%%%%%%%%%%%%%%%%%%%%%%%%%%
(similarly for system $2$), which are in some sense
analogous to spin rotations. $M_{\pm}$ are normalization
factors \cite{normalizace}.
Bell-type experiment then consists of two ``rotations''
according to the recipe (\ref{operace}) performed by two
possibly space-like separated observers, followed by realistic
yes--no detection performed on each mode.
Each such detection has only two possible outcomes (detector either fire or
it does not). Let us assign them values
$z_i$=0 if the detector (in mode $i$) is quiet and
$z_i$=1 if it clicks.
Then the results $X$ and $Y$ of the local two-mode
measurements (including ``rotations'')
performed by the first and second observer, respectively,
can be expressed as
%%%%%%%%%%%%%%%%%%%%%%%%%%%%%%%%%%%%%%%%%%%%%%%%%%%%%
\begin{eqnarray} \label{observation}
X(\theta) &=& z_{24}(\theta)-z_{25}(\theta), \nonumber \\
Y(\theta) &=& z_{34}(\theta)-z_{35}(\theta).
\end{eqnarray}
%%%%%%%%%%%%%%%%%%%%%%%%%%%%%%%%%%%%%%%%%%%%%%%%%%%%%%
After the experiment is repeated
many times and the two observers
compare their notes, the following quantity
can be estimated
%%%%%%%%%%%%%%%%%%%%%%%%%%%%%%%%%%%%%%%%%%%%%%%%%%%%
\begin{equation} \label{def-B}
B=|C(\theta_1,\theta_2)+C(\theta_1,\theta_2')+C(\theta_1',\theta_2)
-C(\theta_1',\theta_2')|,
\end{equation}
%%%%%%%%%%%%%%%%%%%%%%%%%%%%%%%%%%%%%%%%%%%%%%%%%%%%
where correlation function
$$
C(\theta_1,\theta_2) \equiv
\sum_{j,k} X_j Y_k\, p(X_j,Y_k|\theta_1,\theta_2).
$$
As with the two spin-half particles
the use of (\ref{puvodni}) and (\ref{operace}) in (\ref{def-B})
gives $B_{\max}=2\sqrt{2}$, for optimum set of angles.
This exceeds local realistic
limit $B^{\rm cl}_{\rm max}=2$ \cite{Bell}.
Thus from the point of view of
entanglement and local realism the four-mode state (\ref{puvodni})
is equivalent to the maximally entangled state of two
spin-half particles.
This is, of course, consequence of the fact that
the ``cat'' state (\ref{puvodni}) is nothing else than
representation of the maximally entangled spin state in a larger
space -- in the space of four harmonic oscillators.

While spin of e.g.\ neutron can be
easily rotated in magnetic field, it is very difficult to realize
the corresponding unitary transformation (\ref{operace}). Notice
that although the device in Fig.~\ref{schema} does a similar job
(it creates superpositions from the input product states of a
similar sort) its action is not unitary due to the postselection
involved.
This would represent a loophole in the test of local realism when
used in place of the unitary transformation (\ref{operace}).

\section{Decoherence}

So far we assumed that the entangling device was ideal.
This seems natural if fundamental aspects
of {\em thought} device are discussed.
If one, however, seriously thought of experimental implementation
of the entangling device in Fig.~\ref{schema}, such assumption would
clearly be foolish, and various imperfections and unavoidable
losses would have to be considered.

The entangling device can be divided into two parts, the one-photon
MZ interferometer and coherent-state MZ interferometers, with
grossly different sensitivity to imperfections and losses. This can
be illustrated on the simple case of losses modeled e.g.\ by the
presence of an auxiliary beam splitter in the paths of modes
23, 33, and 12.
In the latter case one can always
balance the one-photon interferometer by introducing the same
amount of losses in mode 13, recovering the ideal output
state (\ref{puvodni}) at the expense of decreasing generation rate
\cite{martin}. In such a case, if the photon is detected behind the
interferometer, no information on its path is available and
entanglement in the whole state is kept (if the losses were not
balanced, one path would be more probable).
In contrast to it, a similar compensation of losses in
the coherent-state interferometers cannot help to save the entanglement.
It is not difficult to see why.
The fractions of the
strong coherent beams reflected out of the coherent-state
interferometer always carry a good deal of which-way information
about the photon in the one-photon MZ interferometer,
that can be extracted, e.g., by means of the phase measurement
performed on them. Needles to say, in the
limit of large $\alpha$, when the phase of the reflected beam becomes
sharply defined, the reflected beams carry perfect which-way information
about the photon propagating through the one-photon interferometer;
this degrades the output state (\ref{puvodni}) to the mixed state
(\ref{classic}).
Another likely causes of the loss of  entanglement of the output state
are decoherence effects due to entangling the degrees of freedom of
the quantum system with environment.
First, let us discuss the decoherence in the one-photon
Mach-Zehnder interferometer consisting of modes $12$ and $13$.
We will assume the following simple (but rather general)
model of decoherence,
%%%%%%%%%%%%%%%%%%%%%%%%%%%%%%%%%%%%%%%%%%%%%%%%%%%%
\begin{eqnarray} \label{which-way}
|0\rangle_{12}|g\rangle_{\rm env} &\rightarrow&
|0\rangle_{12}|e_1\rangle_{\rm env},\nonumber \\
|1\rangle_{12}|g\rangle_{\rm env} &\rightarrow&
|1\rangle_{12}|e_2\rangle_{\rm env},
\end{eqnarray}
%%%%%%%%%%%%%%%%%%%%%%%%%%%%%%%%%%%%%%%%%%%%%%%%%%%%
where $|g\rangle_{\rm env}$, $|e_1\rangle_{\rm env}$ and
$|e_2\rangle_{\rm env}$ are three possibly nonorthogonal
states of environment.
Denoting $a$ the overlap
$a=|\langle e_1|e_2\rangle|$,
the mutual information of the output state for
large $|\alpha|$ reads,
%%%%%%%%%%%%%%%%%%%%%%%%%%%%%%%%%%%%%%%%%%%%%%%%%%%%
\begin{eqnarray} \label{dS-1}
I&\approx&2\ln 2+\frac{1+a}{2}\ln\left(\frac{1+a}{2}\right)
\nonumber \\
&&+\frac{1-a}{2}\ln\left(\frac{1-a}{2}\right),\quad
|\alpha|\gg 1.
\end{eqnarray}
%%%%%%%%%%%%%%%%%%%%%%%%%%%%%%%%%%%%%%%%%%%%%%%%%%%%
Similarly, the maximum of the Bell correlation function
is reduced to
%%%%%%%%%%%%%%%%%%%%%%%%%%%%%%%%%%%%%%%%%%%%%%%%%%%%
\begin{equation} \label{B-1}
B_{\max}\approx\sqrt{2}(1+a),\quad |\alpha|\gg 1.
\end{equation}
%%%%%%%%%%%%%%%%%%%%%%%%%%%%%%%%%%%%%%%%%%%%%%%%%%%%
For exact formulas see Appendix.

Note that $a^2$ equals the Ivanovics-Dieks-Peres
lower bound \cite{error-free}
on the probability of the occurrence of an inconclusive
result for error-free discrimination between
the two non-orthogonal states of environment and therefore
also between the states $|0\rangle_{12}$ and $|1\rangle_{12}$.
Hence, one can say that as the amount of principally accessible
which-way information about the photon in the one-photon MZ
interferometer increases, the entanglement and non-classical
character of the output state of the entangling machine become
gradually destroyed. Eqs.~(\ref{dS-1}) and (\ref{B-1}) should
be compared with the corresponding formulas for
entangling machine with (balanced) losses present in the coherent state
MZ interferometers. We will model the losses by
placing four auxiliary beam splitters with the same reflectivity
$R$ into $22$, $23$, $32$ and $33$ paths.
The resulting asymptotic entanglement and maximum
of the Bell correlation  function are
%%%%%%%%%%%%%%%%%%%%%%%%%%%%%%%%%%%%%%%%%%%%%%%%%%%%
\begin{eqnarray} \label{dS-alpha}
I&\approx&2\ln 2+\frac{1+e^{-2R|\alpha|^2}}{2}
\ln\left[\frac{1+\exp(-2R|\alpha|^2)}{2}\right] \nonumber\\
&+& \frac{1-e^{-2R|\alpha|^2}}{2}
\ln\left[\frac{1-\exp(-2R|\alpha|^2)}{2}\right],\quad
|\alpha|\gg1,
\end{eqnarray}
%%%%%%%%%%%%%%%%%%%%%%%%%%%%%%%%%%%%%%%%%%%%%%%%%%%%
and
%%%%%%%%%%%%%%%%%%%%%%%%%%%%%%%%%%%%%%%%%%%%%%%%%%%%
\begin{equation} \label{B-alpha}
B_{\max}\approx\sqrt{2}\left(1+e^{-2R|\alpha|^2}\right),\quad
|\alpha|\gg1.
\end{equation}
%%%%%%%%%%%%%%%%%%%%%%%%%%%%%%%%%%%%%%%%%%%%%%%%%%%%
Notice that Eqs.~(\ref{dS-alpha}) and (\ref{B-alpha})
can be obtained from Eqs.~(\ref{dS-1}) and (\ref{B-1})
simply by substitution
%%%%%%%%%%%%%%%%%%%%%%%%%%%%%%%%%%%%%%%%%%%%%%%%%%%%%
\begin{equation} \label{substit}
a^2\rightarrow e^{-4R|\alpha|^2}.
\end{equation}
%%%%%%%%%%%%%%%%%%%%%%%%%%%%%%%%%%%%%%%%%%%%%%%%%%%%%
This can again be interpreted in terms of available
which-way information. Now, the which-way information
is gained by discriminating between (non-orthogonal)
states  $|\alpha\sqrt{R/2}\rangle$
and $|-\alpha\sqrt{R/2}\rangle$ of the beams reflected
out of paths $23$ and $33$.
The optimum error-free discrimination is
done by mixing the beams with the reference coherent beams
$|i\alpha\sqrt{R/2}\alpha\rangle$
at two mixing beam splitters \cite{alpha-error-free}.
The probability of the inconclusive result:
``no photons detected after the mixing at both outputs'',
is  $P_{\rm inconcl.}=\exp(-2R|\alpha|^2)$ in either case.
The path of the photon is revealed by getting
at least one conclusive result out of the two ideal error-free
measurements in the right and left part of the
entangling apparatus.
The probability of the unfortunate event of
getting two inconclusive results
is $P^{\rm total}_{\rm inconcl.}=P_{\rm inconcl.}^2=
\exp(-4R|\alpha|^2)$, which is just the right hand side of
Eq.~(\ref{substit}).
This suggests that the loss of entanglement and non-locality
caused by both effects, the decoherence in
the one-photon interferometer and losses or decoherence in
the coherent-state interferometers, have common
information-theoretical origin.

Although Eqs. (\ref{dS-alpha})--(\ref{B-alpha}) and
(\ref{dS-1})--(\ref{B-1}) have the same functional dependence on the
amount of available which-way information, their implications
for the feasibility of the generation of the ``macroscopic-cat''
state (\ref{puvodni})  dramatically differ.
The decoherence in the one-photon MZ interferometer is not
a key limiting factor for engineering such superpositions
for Eqs.~(\ref{dS-1}) and (\ref{B-1}) does not depend
on the intensity of the input coherent state $|\alpha|^2$.
In contrast to this, the relative amount of losses in the coherent-state
interferometers which can be tolerated, exponentially decreases with
increasing input intensity, see Eqs.~(\ref{dS-alpha})
and (\ref{B-alpha}). This precludes the entangling of
arbitrarily separated coherent states.

\begin{figure}
\psfig{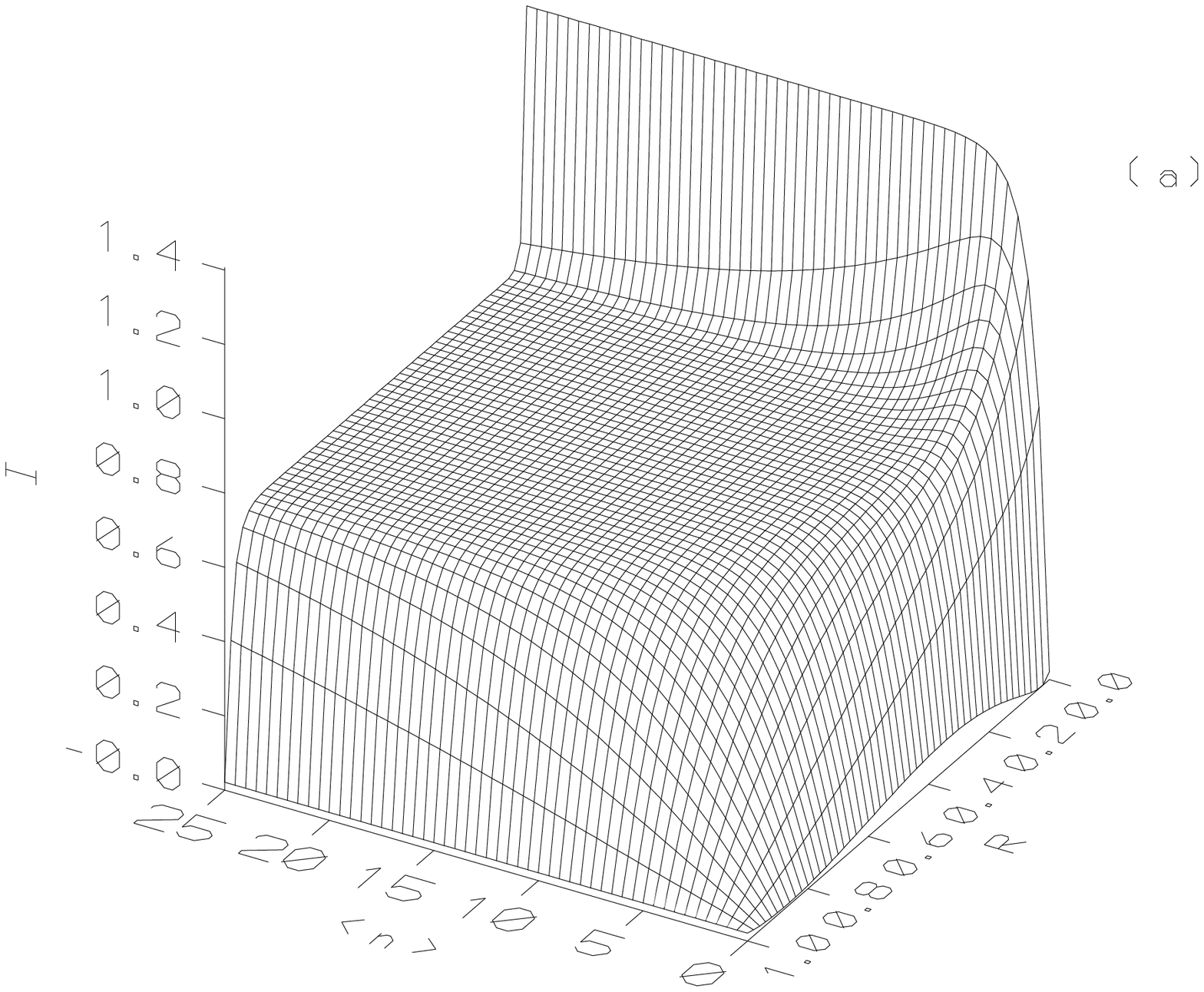}{\hsize}
\psfig{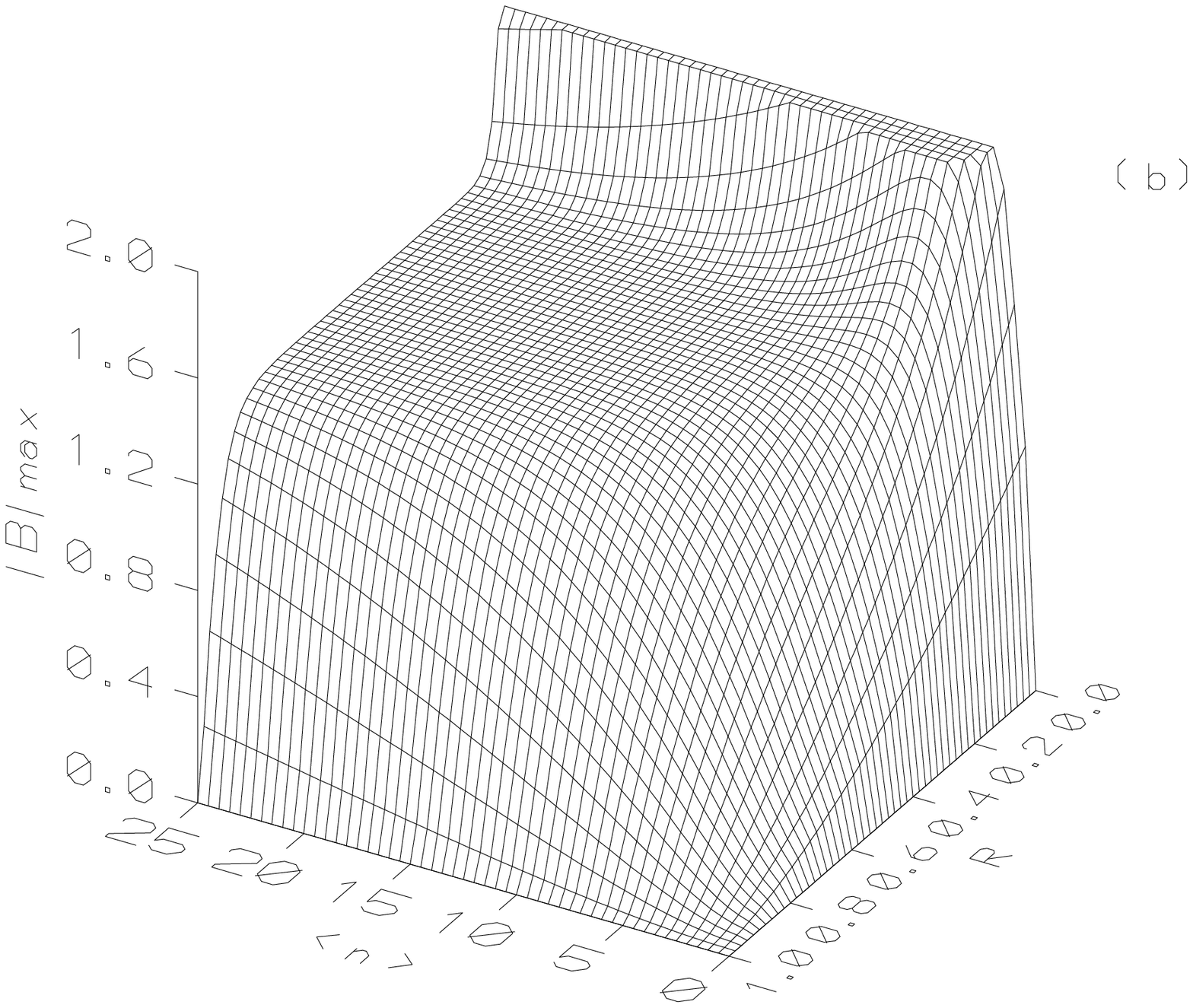}{\hsize}
\vspace*{10mm}
\caption{Mutual information $I_{-}$ (a) and the maximum value of Bell
correlation function $(B_{-})_{\rm max}$ (b) are shown
for different input mean numbers of photons
$\langle n\rangle\equiv|\alpha|^2$
and reflectivities $R$ of the beam splitters simulating losses.}
\label{fig-nonclas}
\end{figure}

Forgetting about macroscopical separability, the presented
device can still be a useful tool for generating interesting
non-classical states. This is illustrated in Fig.~\ref{fig-nonclas}.
The degree of entanglement $I$ in the presence of
losses is shown in Fig.~\ref{fig-nonclas}(a). Notice that
for large $|\alpha|$ the entanglement of the output state
is extremely sensitive to the presence of losses. With
increasing amount of losses the output state becomes mixed state
and the correlations between output modes become explainable classically
(large flat area). For still larger $|\alpha|$, all output modes
get depleted and $I$ drops to zero.
Two scales of sensitivity to disturbances are involved;
one corresponds to the vanishing of nondiagonal elements of
density matrix, the other corresponds to the loss of orthogonality
between correlated output states.
It can be seen that generation of output states having non-classical
correlations is possible for not very large input intensities.

Similar discussion holds also for the maximum attainable value
of Bell correlation function $B_{\rm max}$, see
Fig.~\ref{fig-nonclas}(b). Here part of the graph
lying above the local-realistic limit $B=2$
has been cut off. Resulting upper small flat area adjacent to plane
$R=0$ indicates the range of parameters for which the generation
of nonclassical states is possible. For realistic experimental
setups one again has constraint on the maximum allowed input
intensities.

It follows that experimental generation of nonclassical states
using the proposed device is possible for not very large
intensities $|\alpha|^2$. Such states are interesting from the
point of view of possible interesting experiments on quantum
nonlocality, or experiments utilizing quantum
entanglement, but they are far from being entangled
``macroscopic'' states.

\section{Conclusion}

We have shown that the ``entangling'' apparatus proposed in
Ref.~\cite{howell} could never produce entangled pairs of coherent
states. We have designed a device that can do it. This device
also uses Kerr nonlinear medium which helps to extend the one-photon
non-separable superposition to the four-mode entangled superposition of
strong coherent fields. The new important point is a postselection
based on interferometric measurement on the one-photon subsystem.
This erases which-way information that had prevented the creation of desired
entangled state. We have proved that the states prepared by our
prescription can violate Bell-like inequalities. We have also studied to
which extend loses and decoherence can degrade the produced state. This
is important with respect to potential experimental realization.
Unfortunately, the preparation procedure is very sensitive to
decoherence and especially to losses in the strong-field interferometers.
However, some set of realistic values in parameter space still exists for
which entropy of states exceeds classical level and even Bell inequality
can be violated.

\section*{Acknowledgment}
This work was supported under the project LN00A015 and Research Plan
CEZ:J14 ``Wave and Particle Optics'' of the Ministry
of Education of the Czech Republic, and project
19982003012 of the Czech National Security Authority.

\appendix

\section{Some exact formulae}

Explicit formulae for mutual information (\ref{entrop})
and Bell-factor (\ref{def-B}) of the system in
states (\ref{bily}) ($+$) and (\ref{cerny}) ($-$)
are as follows
\begin{equation}
I_{\pm}=p_{1\pm}\ln p_{1\pm}+p_{2\pm}\ln p_{2\pm}-2(p^{r}_{1\pm}\ln
p^{r}_{1\pm}+p^{r}_{2\pm}\ln p^{r}_{2\pm}),
\end{equation}
\begin{eqnarray}
B_{\pm}=|C_{\pm}(\theta_{I},\theta_{II})+C_{\pm}(\theta_{I},\theta^{'}_{II})+
\nonumber\\ +C_{\pm}(\theta^{'}_{I},\theta_{II})-C_{\pm}
(\theta^{'}_{I},\theta^{'}_{II})|.
\end{eqnarray}
Here
\begin{eqnarray}
p_{1\pm}&=&\frac{(1\pm d)(1+\mu^{2})}{2(1\pm\mu^{2} d)},\nonumber\\
p_{2\pm}&=&\frac{(1\mp d)(1-\mu^{2})}{2(1\pm\mu^{2} d)},\nonumber\\
p^{r}_{1\pm}&=&\frac{(1\pm\mu d)(1+\mu)}{2(1\pm\mu d)},\nonumber\\
p^{r}_{2\pm}&=&\frac{(1\mp\mu d)(1-\mu)}{2(1\pm\mu d)},
\end{eqnarray}
\begin{eqnarray}
C_{\pm}(\theta_{I},\theta_{II})&=&N_{\pm}^{2}\left((N_{1}+N_{2})
\cos(2\theta_{I})\cos(2\theta_{II})\mp\right.\nonumber\\
& &\left.\mp 2dN_{3}\sin(2\theta_{I})\sin(2\theta_{II})\right),
\end{eqnarray}
and
\begin{eqnarray}
N_{1}&=& \frac{1}{(1+\mu\sin(2\theta_{I}))(1-\mu\sin(2\theta_{II}))},\nonumber\\
N_{2}&=& \frac{1}{(1-\mu\sin(2\theta_{I}))(1+\mu\sin(2\theta_{II}))},\nonumber\\
N_{3}&=&
\frac{1}{\sqrt{(1-\mu\sin^{2}(2\theta_{I}))(1-\mu\sin^{2}(2\theta_{II}))}}.
\end{eqnarray}
Factor $\mu$ accounts for various disturbances.
In the case of decoherence in the one-photon interferometer
it reads
\begin{equation}
\mu=\exp(-|\alpha|^{2}),\hspace{0.5cm} d=\sqrt{1-|\langle
e_{1}|e_{2}\rangle|^{2}}.
\end{equation}
In the case of balanced losses modeled by beam splitters present
in the coherent-state interferometers one has
\begin{equation}
\mu=\exp(-T|\alpha|^{2}),\hspace{0.5cm} d=\exp(-2R|\alpha|^{2}).
\end{equation}
The maximum of $|B_{-}|$ occurs for angles
\begin{equation}
\theta_{I}=0,\theta^{'}_{I}=\frac{\pi}{4},\theta_{II}=\frac{\pi}{8},
\theta^{'}_{II}=-\frac{\pi}{8}.
\end{equation}

\vspace{1cm}

\vspace{1cm}


\begin{thebibliography}{9}


\bibitem{howell}
J.C. Howell and J.A.Yeazell, Phys. Rev. A {\bf 62}, 012102 (2000).

\bibitem{macro}
 B.\,C.~Sanders, and G.\,J.~Milburn, Phys.\ Rev.\ A~{\bf 39}, 694
 (1989);
 B.\,C.~Sanders, Phys.\ Rev.\ A~{\bf 45}, 6811 (1992); {\bf 46}, 2966
 (1992);
 B.~Wielinga and B.\,C.~Sanders, J.\ Mod.\ Opt. {\bf 40}, 192 (1993);
 A.~Mann, B.\,C.~Sanders, and W.\,J.~Munro, Phys.\ Rev.\ A~{\bf 51}, 989
 (1995);
 B.\,C.~Sanders, K.\,S.~Lee, M.\,S.~Kim, Phys.\ Rev.\ A~{\bf 52}, 735
 (1995);
 D.\,A.~Rice, and B.\,C.~Sanders, Quantum.\ Semiclass.\ Opt.\ {\bf 10},
 L41 (1998);
 B.\,C.~Sanders, and D.\, A.~Rice, Opt.\ Quantum Electron.\ {\bf 31},
 525 (1999);
 B.\,C.~Sanders, and D.\, A.~Rice, Phys.\ Rev.\ A~{\bf 61}, 013805
 (2000);
 D.\,A.~Rice, G.~Jaeger, and B.\,C.~Sanders, Phys.\ Rev.\ A~{\bf 62}, 012101
 (2000);


\bibitem{normalizace}
$N_{\pm}=[2\pm2\exp(-2|\alpha|^2]^{(-1/2)}$; $M_{\pm}=
[1\pm\exp(-|\alpha|^2)\sin 2\theta]^{(-1/2)}$.


\bibitem{eraser}
 M.~Hillery and M.\,O.~Scully, in {\em Quantum Optics, Experimental
   Gravitation, and Measurement Theory}, eds. P.~Meystre and
   M.\,O.~Scully (Plenum, New York, 1983), p.~65;
 M.\,O.~Scully, B.\,G.~Englert, and H.~Walther, Nature {\bf 351}, 111
   (1991);
 P.\,G.~Kwiat, A.\,M.~Steinberg, and R.\,Y.~Chiao, Phys.\ Rev.\
   A~{\bf 45}, 7729 (1992);
 P.\,G.~Kwiat, M.~Aerphraim, A.\,M.~Steinberg, and R.\,Y.~Chiao, Phys.\ Rev.\
   A~{\bf 49}, 61 (1994).

 \bibitem{barnett} S.M.Barnett and S.J.D.Phoenix, Phys. Rev. A
{\bf 40}, 2404 (1989).


\bibitem{correl} The diagonal representation of the
density matrix of two correlated subsystems
displaying maximum classically allowed correlations is of the form
$\hat{\rho}=\sum_i p(a_i,b_i) |a_i\rangle_1|b_i\rangle_2
\langle b_i|_{2}\langle a_i|_1$.
Then one gets $S=S_1=S_2$.

\bibitem{EPR}
A.~Einstein, B.~Podolsky, N.~Rosen, Phys.\ Rev.\ {\bf 47}, 777 (1935);
D.~Bohm, Phys.\ Rev.\ {\bf 85}, 166 (1952);
N.~Bohr, Nature {\bf 136}, 65 (1935).

\bibitem{Bell}
J.\,S.~Bell, Physics {\bf 1}, 195 (1964);
J.\,F.~Clauser, M.\,A.~Horne, A.~Shimony, R.\,A.~Holt, Phys.\ Rev.\
   Lett. {\bf 23}, 880 (1969).

\bibitem{martin}
  M.~Hendrych, M.~Du\v{s}ek, O.~Haderka,
  Acta Physica Slovaka {\bf 46}, 393 (1996).

\bibitem{error-free}  I.D. Ivanovic, Phys. Lett. A {\bf 123},
257 (1987); D. Dieks, Phys. Lett. A {126}, 303 (1998);
A. Peres, Phys. Lett. A {\bf 128}, 19 (1988).

\bibitem{alpha-error-free} B. Huttner, N. Imoto, N. Gisin, and
T. Mor, Phys. Rev. A {51}, 1863 (1995); L.S. Phillips, S.M.Barnett,
and D.T. Pegg, Phys. Rev. A {\bf 58}, 3259 (1998).

\end{thebibliography}
\end{document}